\documentclass[a4paper]{jpconf}
\usepackage{graphicx}
\usepackage{amsmath}
\usepackage[utf8]{inputenc}
\def\@evenhead{\underline{\protect\parbox{\textwidth}{\bf\boldmath\protect\rule[-0.2cm]{0pt}{0.2cm}{\Large\thepage}\hfill\leftmark\hfill~}}}%
\def\@oddhead{\underline{\protect\parbox{\textwidth}{\bf\boldmath\protect\rule[-0.2cm]{0pt}{0.2cm}~\hfill\rightmark\hfill{\Large\thepage}}}}%

\newcommand{\mytimes}{\kern-0.12em \times \kern-0.2em}
\newcommand{\mysim}{\sim \kern-0.2em}

\newcommand{\eVdist}{\kern-0.06667em}

\newcommand{\Pom}{{\text{I\kern-0.12em P}}}

\newcommand{\gevv}     {\mbox{${\rm Ge\kern -0.1em V}^{2}$}}

\newcommand{\lw}[1]{\smash{\lower1.8ex\hbox{#1}}}

\catcode`\@=11 
\newcommand{\parenbar}{\mathpalette\p@renb@r}
\def\p@renb@r#1#2{\vbox{%
  \ifx#1\scriptscriptstyle \dimen@.7em\dimen@ii.2em\else
  \ifx#1\scriptstyle \dimen@.8em\dimen@ii.25em\else
  \dimen@1em\dimen@ii.4em\fi\fi \offinterlineskip
  \ialign{\hfill##\hfill\cr
    \vbox{\hrule width\dimen@ii}\cr
    \noalign{\vskip-.3ex}%
    \hbox to\dimen@{$\mathchar300\hfil\mathchar301$}\cr
    \noalign{\vskip-.3ex}%
    $#1#2$\cr}}}
\catcode`\@=12 

\newbox\struttbox
\setbox\struttbox=\hbox{\vrule height1.65ex depth.485ex width0pt}
\def\strutt{\relax\ifmmode\copy\struttbox\else\unhcopy\struttbox\fi}
\def\stru#1#2{\relax\ifmmode\hbox{\vrule height#1 depth#2 width0pt}
\else\vrule height#1 depth#2 width0pt\fi}

\begin{document}
\title{Neutrino Astronomy with the IceCube Observatory}

\author{A Kappes$^{a,b}$ for the IceCube Collaboration\footnote{http://www.icecube.wisc.edu}}

\address{$^a$Humboldt-Universität zu Berlin, Institut für Physik, Newtonstr. 15, 12489 Berlin, Germany\\$^b$DESY, Platanenallee 6, 15738 Zeuthen, Germany}

\ead{kappes@desy.de}

\begin{abstract}
IceCube is the first representative of the km$^3$ class of neutrino telescopes and currently the most sensitive detector to high-energy neutrinos. Its main mission is to search for Galactic and extragalactic sources of high-energy neutrinos, but it is also an excellent detector for the investigation of a variety of other highly topical astrophysics and particle physics topics like supernovae, dark matter and neutrino oscillations. After an introduction to neutrino astronomy and neutrino telescopes, this article presents a selection of latest results from the IceCube neutrino detector with respect to searches for cosmic high-energy neutrino sources.
\end{abstract}

\section{Introduction}
On August 7, 2012, we celebrated the $100^\mathrm{th}$ anniversary of the discovery of the cosmic radiation by Victor Hess for which he received the Nobel price in 1936\footnote{As not uncommon in scientific history, there were  several other scientists who came to similar conclusions with less precise or direct experiments earlier like Gockel 1909 and Pacini 1910, or  confirmed the measurements with much improved systematics shortly after like Kohlhörster 1913.}. First, people including Hess thought that this radiation consisted of gamma rays as only those were known to have the required penetration power. Later it was discovered that the cosmic radiation mainly consists of protons and to a smaller extend of heavier nuclei but the name \emph{cosmic radiation} survived. In the last 100 years we have learned a great deal more about their composition, energy spectra etc. For example, we know from measurements that they span an energy range from a few GeV up to $10^{20}$\,eV. Those below about $10^{15}$\,eV are thought to be of Galactic origin whereas those above about $10^{18}$\,eV are thought to be produced by extragalactic sources. However, the exact sources, even source types, are still unknown, as the charged particles of the cosmic radiation are deflected in Galactic and extragalactic magnetic fields and hence loose all their directional information concerning their origin except maybe at the highest energies. Finding the sources of cosmic radiation requires therefore neutral (and stable) messenger particles of which the standard model of particle physics contains only two, the photon and the neutrino. With the advances in gamma-ray astronomy in the last decades people thought that soon the mystery of the origin of cosmic rays would be solved. However, even after the discovery of over 100 gamma-ray sources in the TeV range with  imaging Cherenkov telescopes during the last decade, smoking-gun evidence for the sources of the Galactic cosmic rays is still missing and the origin of the UHE cosmic rays remains a complete mystery. 

In contrast, the observation of high-energy neutrinos from an object or region unambiguously identifies it as a source of high-energy cosmic rays, which has been one of the main drivers for the development and construction of so-called neutrino telescopes over the last decades. Unfortunately however, early predictions of neutrino fluxes from various sources turned out to be too optimistic by an order of magnitude and today detectors of at least km$^3$ size are considered necessary for the detection of the first cosmic high-energy neutrino source (see e.g.\ \cite{apj:656:870,app:34:778}). The IceCube observatory, consisting of a neutrino telescope buried in the deep ice beneath the South Pole and a cosmic ray detector on top\footnote{For more information on this detector and cosmic-ray physics with IceCube see contribution from H. Kolanoski in these proceedings.}, is the first representative of this km$^3$ class of neutrino telescopes\footnote{A multi-km$^3$ neutrino telescope in the Mediterranean Sea, KM3NeT, is being planned and preparations for a first construction phase have started.}  and was completed in December 2010 (data was also taken with the partially completed detector from 2006 on; this data is referred to by IC\emph{XX} with \emph{XX} giving the number of installed strings). The neutrino-detector part consists of 86 vertical strings each equipped with 60 10-inch photomultipliers housed in pressure vessels in depths between 1450\,m and 2450\,m. For a more detailed description of the detector see \cite{app:26:155}. 

Neutrinos are reconstructed by detecting the arrival time and intensity of Cherenkov light from charged secondary particles, which are produced in interactions of the neutrinos with the nuclei in the surrounding ice or the bedrock below the detector. Two basic event topologies can be distinguished: track-like patterns of detected Cherenkov light (hits) which originate from muons produced in charged-current interactions of muon neutrinos (muon channel); and spherical hit patterns which originate from the hadronic cascade at the vertex of neutrino interactions or the electromagnetic cascade of electrons from charged current interactions of electron neutrinos (cascade channel). If the charged current interaction happens inside the detector or in case of charged current tau-neutrino interactions, these two topologies overlap which complicates the reconstruction. The direction of elongated muon tracks can be reconstructed significantly better than that of cascades reaching $\sim 0.5^\circ$ at high energies. At the relevant energies, the neutrino is approximately collinear with the muon and, hence, the muon channel is the prime channel for the search for point-like sources of cosmic neutrinos. On the other hand, cascades deposit all of their energy inside the detector and therefore allow for a much better energy reconstruction with a resolution of a few 10\%. 

\begin{figure}
\includegraphics[width=8.2cm]{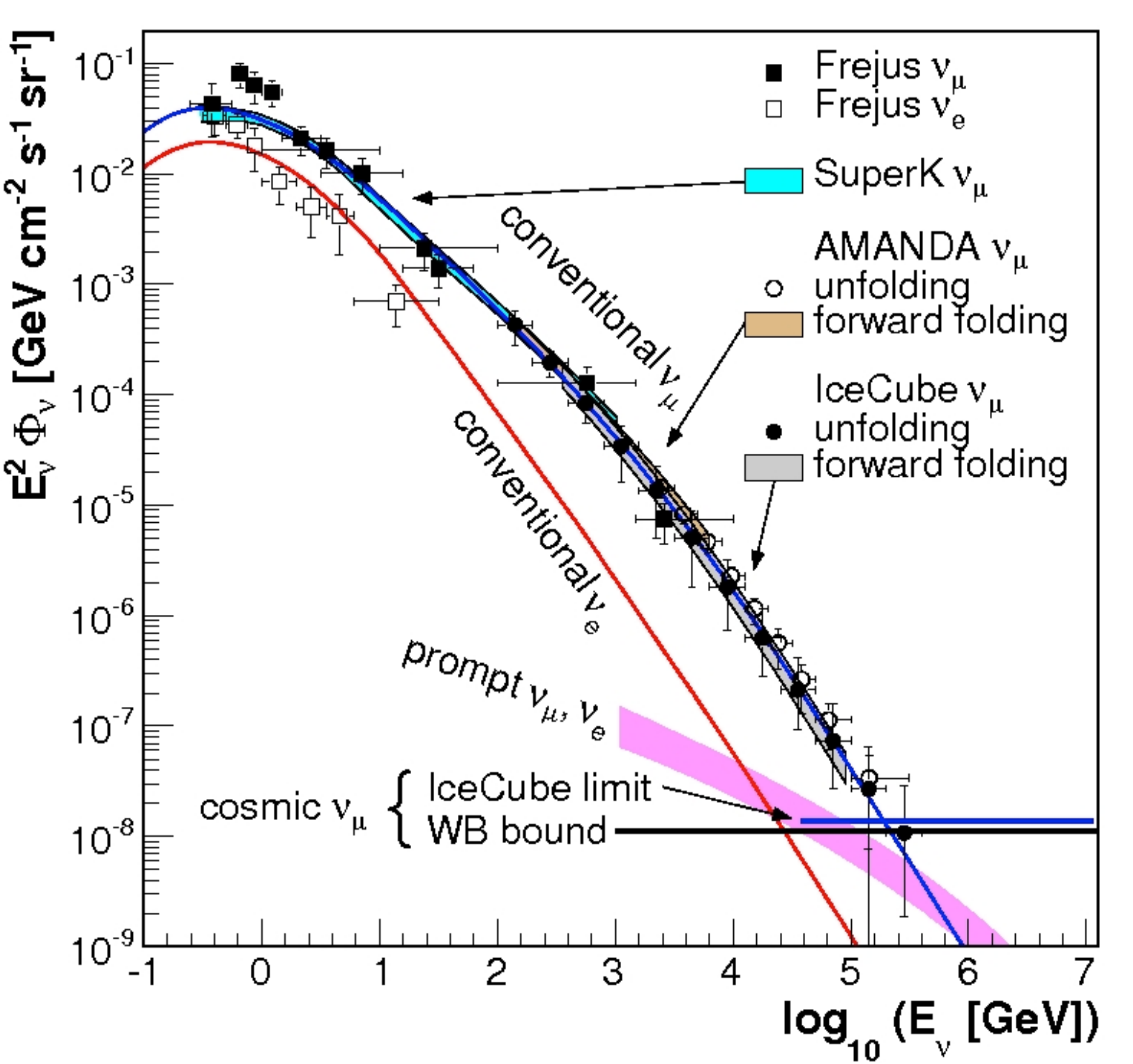}\hfill%
\begin{minipage}[b]{7cm}
\caption{\label{fig:atm_nu}Atmospheric muon and electron neutrino spectrum as function of energy. The open and filled symbols represent measurements of various detectors (IceCube results are from the 40-string configuration). Curved lines are theoretical predictions of atmospheric fluxes (the (magenta) band for the prompt flux indicates the theoretical uncertainty) and the two horizontal lines are the Waxman-Bahcall bound and the IceCube upper limit (59-string configuration) for cosmic muon neutrinos. }
\end{minipage}
\end{figure}

\section{Atmospheric muons and neutrinos}
The main background for the detection of cosmic neutrinos originates from the interaction of cosmic rays with the Earth's atmosphere producing muons and neutrinos. Atmospheric muons are commonly suppressed by observing the northern hemisphere where they are absorbed by the Earth. However, for energies above $\sim$100\,TeV the Earth starts to become opaque also for neutrinos which makes the southern hemisphere the only part of the sky from which neutrinos with EeV energies and above can be observed. Fortunately, at these energies the atmospheric muon background is negligible. 

Atmospheric neutrinos are an irreducible background for all searches for cosmic neutrinos. On the other hand, they are invaluable as a calibration source and also allow to investigate very interesting physics topics like for example neutrino oscillations. Figure~\ref{fig:atm_nu} shows the atmospheric muon and electron neutrino fluxes as measured by several experiments together with theoretical predictions. The \emph{conventional} neutrino flux  \cite{pr:d75:043006} originates from the decay of kaons and pions produced in the cosmic-ray interactions in the atmosphere ($\pi^\pm,K^\pm \rightarrow \mu^\pm + \parenbar{\nu}_\mu$) and the subsequent muon decay. Electron neutrinos are less abundant than muon neutrinos as they are only generated in the muon decay. The ratio decreases with increasing energy as an ever larger number of muons reach the detector before decaying. The \emph{prompt} neutrino flux stems from the decay of charmed particles generated in the same cosmic ray interactions. As these particles are very short-lived they decay without interaction and therefore yield a harder spectrum (about equal flux for muon and electron neutrinos) compared to the conventional neutrino flux. Currently, there exists no measurement of this prompt flux and its normalization has large theoretically uncertainties \cite{pr:d78:043005} .

The measurements are well described by the theoretical predictions of the conventional neutrino flux. The measurements for muon neutrinos cover a broad energy range up to energies of several 100\,TeV observed by IceCube. IceCube's sensitivity for astrophysical neutrinos is well below the Waxman-Bahcall upper bound \cite{pr:d59:023002}. The latest search for astrophysical muon neutrinos with IC-59 yielded a flux limit above the Waxman-Bahcall bound though, because the observed energy spectrum is slightly harder than that predicted for atmospheric neutrinos alone while still being consistent with no contribution from astrophysical neutrinos. Compared to the muon neutrino channel, the energy reach in electron neutrinos is much smaller but first results in the high-energy regime can be expected  from IceCube in the near future. 

\section{Neutrinos from Galactic sources}

\begin{figure}
\includegraphics[width=8.5cm]{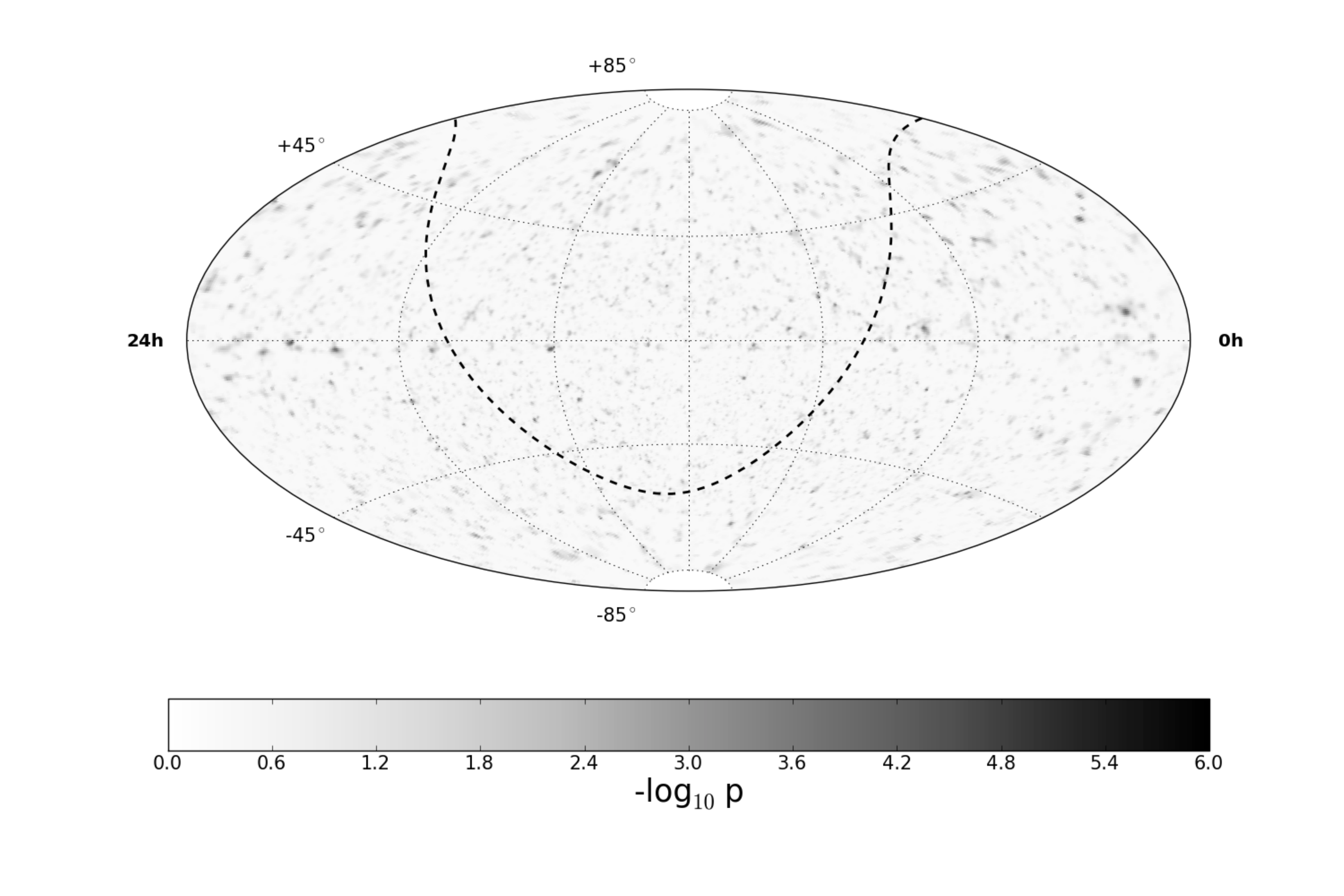}\hfill%
\begin{minipage}[b]{6cm}
\caption{\label{fig:ps_skymap}IceCube significance map for cosmic point-like neutrino sources in equatorial coordinates for the combined IC40+IC59 data set. The color scale indicates the p-value before correction for all-sky trial factors.}
\end{minipage}
\end{figure}

As cosmic neutrino spectra are believed to be harder than the atmospheric one,  the signal to noise ratio decreases with increasing energy for any search for cosmic neutrinos. The supposed cutoff in the Galactic cosmic-ray spectrum around $3\times10^{15}$\,eV translates into a cutoff in the associated neutrino spectrum around 200\,TeV. Hence, for neutrino telescopes the interesting energy range for Galactic sources is the TeV range.

Searches for point-like sources have the advantage that for a particular source they look only at a small portion of the sky (IceCube's angular resolution for muon neutrinos is about $0.5^\circ$) thereby reducing the background of atmospheric neutrinos considerably. Figure~\ref{fig:ps_skymap} shows the significance (p-value) skymap before trial-factor correction in equatorial coordinates for point-like sources obtained from the analysis of the combined IC40 and IC59 data equivalent to about one year of data taking with the full detector. Note that the northern sky is dominated by (atmospheric) neutrinos (about 58000 events) whereas the southern sky mainly consists of atmospheric muons (about 87000 events). For the latter, the energy cuts had to be tightened to allow for a possible cosmic signal to emerge from the background. As a consequence, the sensitivity in the southern sky for neutrinos with an $E^{-2}$ spectrum is in the PeV range whereas in the northern sky it is in the TeV range. After correction for the fact, that the map is evaluated at many positions (trial factor), the probability to obtain a fluctuation at least as high as the highest observed one (located in the southern sky) anywhere on the sky map purely from background fluctuations  (post-trial) is 67\% and hence not significant. The investigation of 43 preselected sources (13 Galactic, 30 extragalactic), which reduces the trial factor considerably compared to the all-sky search,  also yields no significant deviation from the background hypothesis with the lowest p-values lying above 10\% (Cen\,A, PKS\,1454$-$354). 

\begin{figure}
\includegraphics[width=8.5cm]{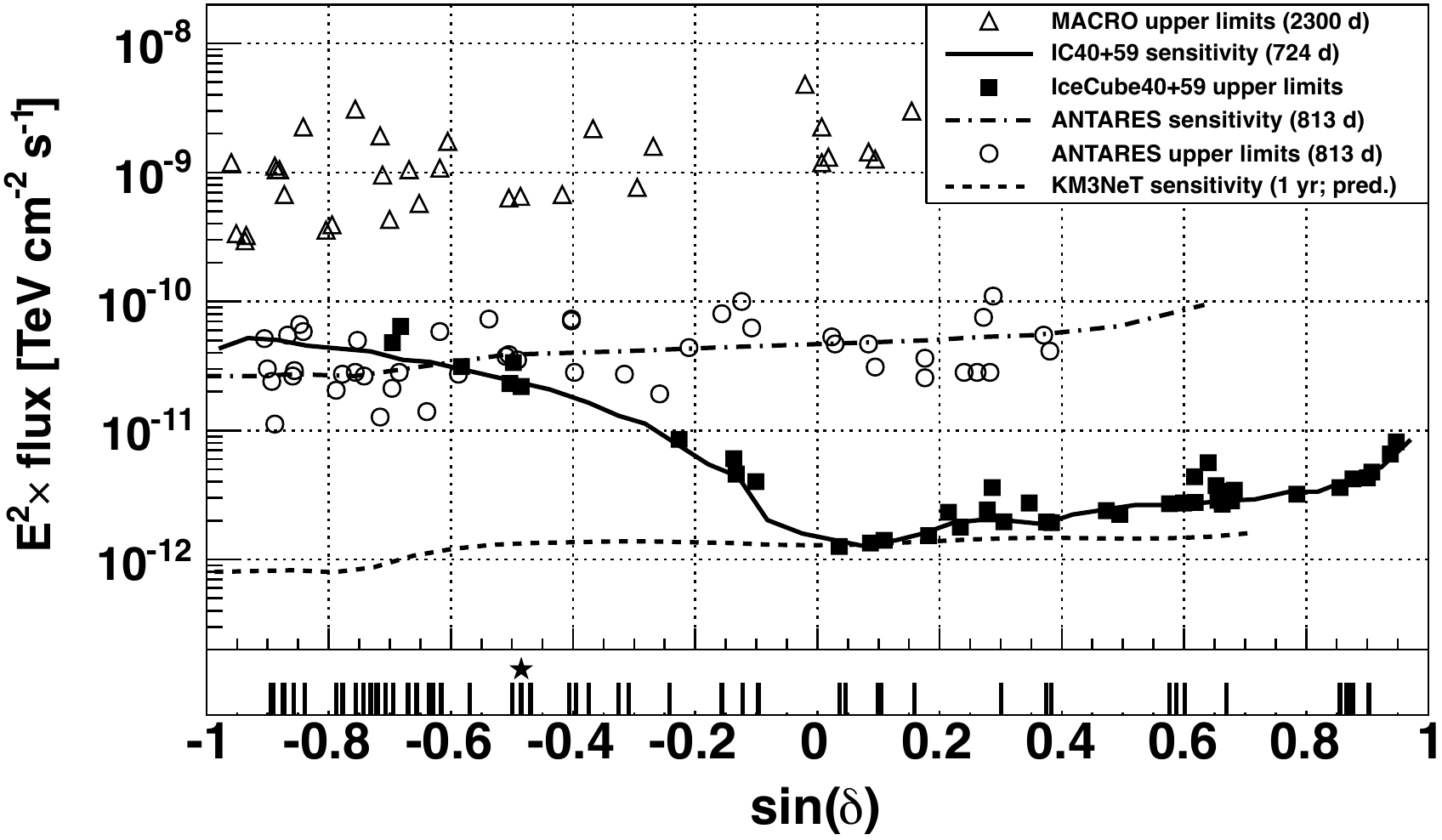}\hfill%
\begin{minipage}[b]{6cm}
\caption{\label{fig:ps_limits}Upper limits (symbols) and sensitivities (lines) (both 90\% CL) for point-like sources with an $E^{-2}$ spectrum as function of declination. The vertical lines at the bottom represent the positions of known Galactic TeV gamma-ray emitters. The star marks the position of the Galactic center. }
\end{minipage}
\end{figure}

The non-observation of a significant deviation from the background hypothesis leads to upper limits which are plotted in Fig.~\ref{fig:ps_limits} (solid squares) as function of declination for a cosmic $E^{-2}$ neutrino flux together with the upper limits (symbols) and sensitivities (lines) from other experiments. The plot illustrates the large improvement in sensitivity of factor 1000 over the last 15 years. Currently, IceCube is the most sensitive operating neutrino telescope which in the northern sky (viewed using Earth as a neutrino filter) starts to approach the \emph{discovery region} below $\sim10^{-12}\,\mathrm{TeV}\,\mathrm{cm}^{-2}\,\mathrm{s^{-1}}$ where according to current calculations \cite{apj:656:870,app:34:778} (Galactic) fluxes are expected to lie. For viewing the southern sky, ANTARES located in the Mediterranean Sea is currently the most sensitive detector\footnote{Note that the sensitivity of IceCube in the  northern sky is in the PeV range whereas for ANTARES it is in the TeV range, and that \emph{Galactic} sources are expected to emit neutrinos in the TeV range.} \cite{arxiv:1207.3105}. However, due to its  $\sim100$ times smaller volume it has a much lower peak sensitivity. Therefore, in order to reach a full high-sensitivity sky coverage, in particular of the inner Galactic plane and the Galactic center,  a (multi-) km$^3$-scale neutrino telescope in the Northern Hemisphere is needed. Such a detector, KM3NeT \cite{km3net:tdr-short:2010}, to be installed in the Mediterranean Sea, is currently in the prototyping phase with funds for a first construction stage available. 

\section{Neutrinos from extragalactic sources}

\begin{figure}[t]
\includegraphics[width=9.2cm]{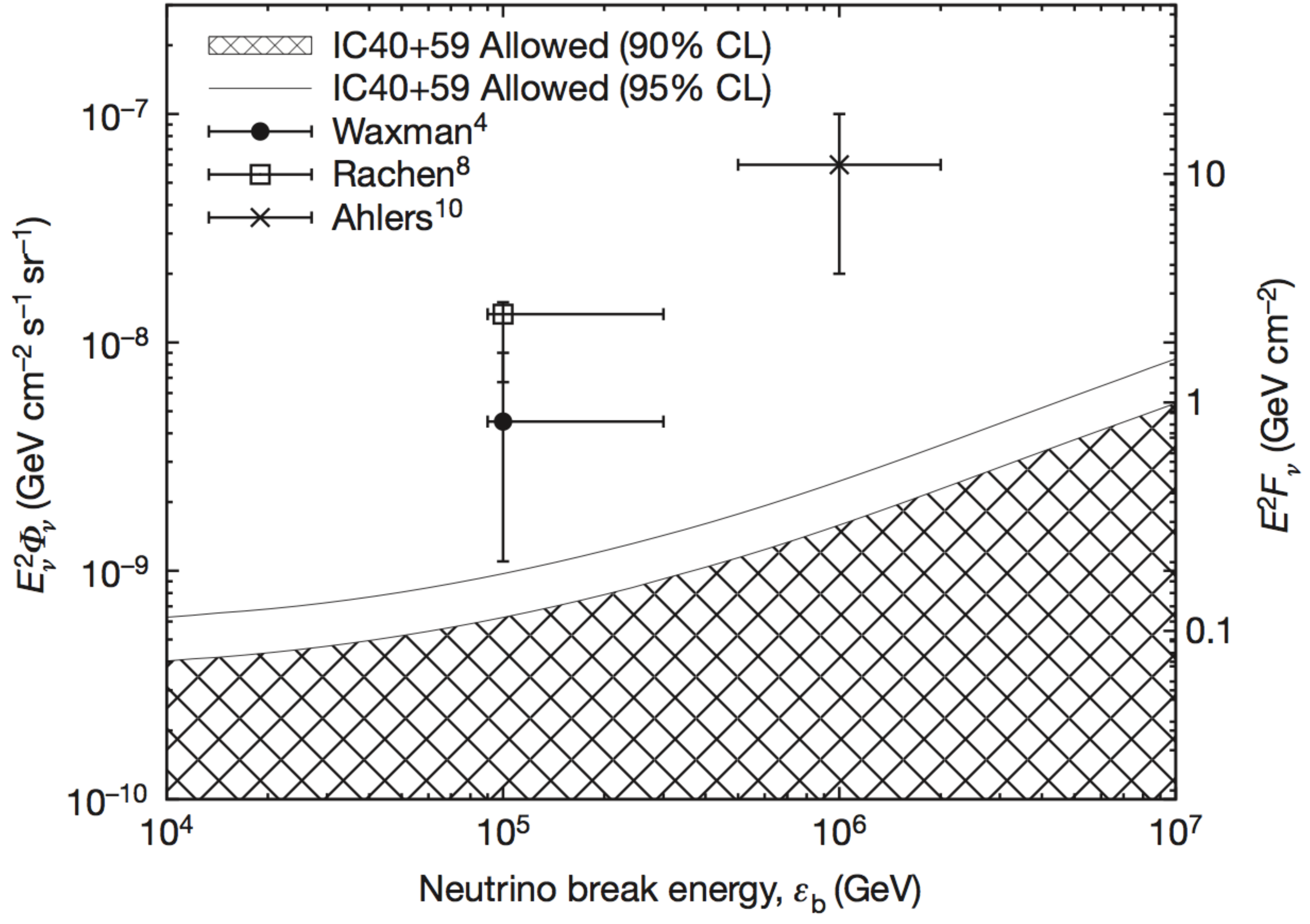}\hfill%
\begin{minipage}[b]{6cm}
\caption{\label{fig:grb_nuescape}Compatibility of some neutrino flux predictions based on cosmic
ray production in GRBs with observations from IceCube. The cross-hatched area shows the 90\% confidence allowed values of the neutrino flux versus the first break energy in the neutrino spectrum in comparison to model predictions with estimated uncertainties (points); the solid line shows the upper bound of the 95\% confidence allowed region. Taken from \cite{na:484:351}. }
\end{minipage}
\end{figure}

The origin of the extragalactic cosmic rays in general and those at ultra-high energies (UHE) above $10^{19}$\,eV in particular are still a mystery. For the latter, currently only two source classes are considered to provide the necessary environment: Active Galactic Nuclei (AGNs) and Gamma-Ray Bursts (GRBs). Due to the particular environment of AGNs, predictions of neutrino fluxes are difficult and have orders of magnitude uncertainties. The source list presented in the previous section contains 30 AGNs where none shows a significant deviation from background in the number of observed events.

Neutrino flux predictions for GRBs are to some extend more straight forward and generally yield double-broken power-law spectra with first and second break energies in the 100\,TeV and PeV range, respectively. However, the models still contain several parameters which are poorly determined leading to significant uncertainties in the predicted fluxes. A quantification of this uncertainty is hindered by the correlation of the different parameters among each other. Analyzing the combined IC40/IC59 dataset \cite{na:484:351} yields no events associated with any of the 225 GRBs which occurred during data taking, whereas the model following \cite{app:20:429} predicts 8.4 events. This corresponds to an upper limit (90\% CL) which is a factor 3.7 below the predictions. It should be noted here, that in addition to the uncertainties in the parameters mentioned above, it was found recently that simplifications in the implemented particle physics processes lead to a significant overestimation of the predicted fluxes in this particular model \cite{prl:108:231101}. On the other hand, models like \cite{app:35:87} which use the observed flux of cosmic rays at UHE to normalize the neutrino flux are more robust against these uncertainties. Figure~\ref{fig:grb_nuescape} shows some of the latter models together with estimated uncertainties and compares them to the parameter range allowed by the IceCube analyses (for more details see \cite{na:484:351}). Though, in view of the uncertainties of the models, it is probably too early to exclude GRBs as sources of extragalactic cosmic rays, continuing negative results from IceCube in the upcoming years will certainly seriously question the role of GRBs as major sources of UHE cosmic rays. It should also be noted that in the case that IceCube established GRBs as major sources of UHE cosmic rays, the latter would have to mainly consist of protons as heavier nuclei are expected to disintegrate in the dense photon environment of GRB jets.

\section{Cosmogenic neutrinos}

\begin{figure}[t]
\begin{minipage}[t]{7.5cm}
	\includegraphics[width=7.5cm]{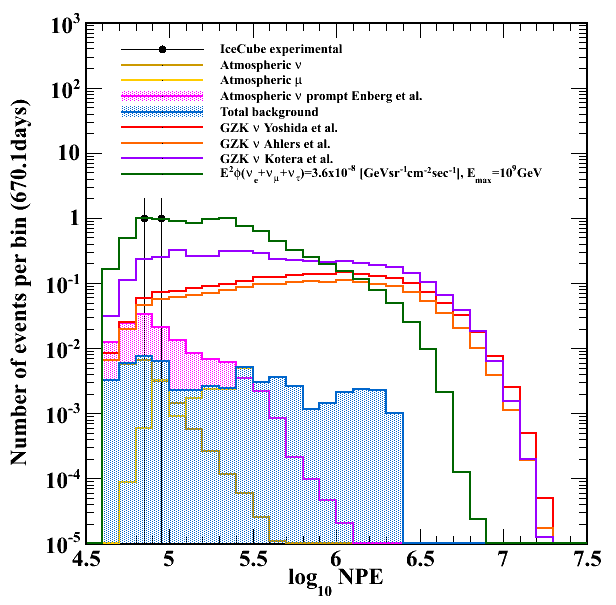}
	\caption{\label{fig:ehe_events}Number of events per bin as function of the number of photo electrons measured in the detector. Points are data, lines represent theoretical predictions of cosmic fluxes and filled histograms are atmospheric background.}
\end{minipage}\hfill%
\begin{minipage}[t]{7.5cm}
	\includegraphics[width=7.5cm,height=7.45cm]{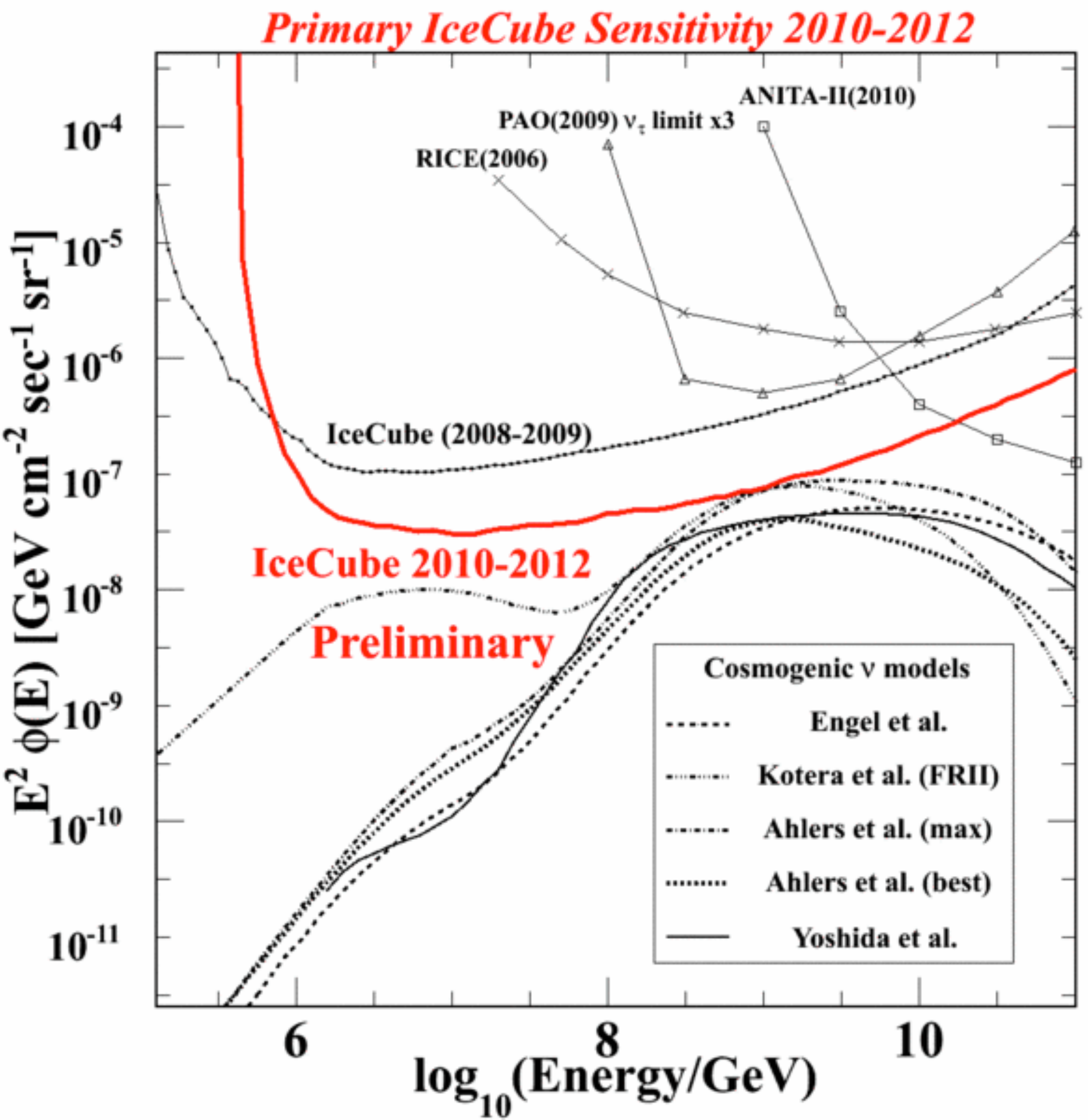}
	\caption{\label{fig:ehe_limit}Differential flux limits (symbols with lines) as function of neutrino energy. For IceCube the bin width is one decade in energy. In addition, several theoretical predictions of cosmogenic fluxes are shown (lines).}
\end{minipage} 
\end{figure}

Under the assumption that the UHE cosmic rays consist (mainly) of protons, a guaranteed source of high-energy neutrinos originates from the interaction of these protons with photons of the microwave background radiation (\emph{cosmogenic neutrinos}, see e.g.\ \cite{app:34:106}). Theoretical uncertainties of the expected flux mainly arise from the source distribution and their evolution with redshift. The combined analysis of IC79 and the first year of IC86 data, tuned to specifically select very-high energy events, finds two events at the lower end of the energy-proxy range (Fig.~\ref{fig:ehe_events}) with 0.06 expected from atmospheric background. This corresponds to a significance of $2.9\,\sigma$ but doesn't include systematic uncertainties and in particular not the unknown contribution from prompt neutrinos. According to common models, the latter would yield an additional background contribution of about 0.13 reducing the significance to about $2.2\,\sigma$. The resulting differential upper limit from the analysis is displayed in Fig.~\ref{fig:ehe_limit}. It is the most stringent one below $10^{19}$\,eV and touches the range of common cosmogenic flux predictions.

\section{Conclusions}
Though all searches for high-energy cosmic neutrinos have been unsuccessful up to now, the large improvements in sensitivity with the IceCube detector already in its first years of operation allows us to enter for the first time into the regions of predicted cosmic neutrino fluxes. Together with first tentative  signs for deviations from the atmospheric background expectations the discovery of the first cosmic neutrinos might be just around the corner. Together with the KM3NeT detector in the Mediterranean Sea, neutrino astronomy might finally kick off and deliver the fruits of decades of hard efforts.

\vspace{2mm}
{\bf\noindent References}
\providecommand{\newblock}{}

\end{document}